\documentclass[a4paper,10pt]{article}
\usepackage[dvips]{graphicx}
\usepackage{amssymb,amsmath}
\numberwithin{equation}{section}

\begin{document}

 \def\gsim{ \lower .75ex \hbox{$\sim$} \llap{\raise .27ex \hbox{$>$}} }
 \def\lsim{ \lower .75ex \hbox{$\sim$} \llap{\raise .27ex \hbox{$<$}} }



 \title{\Large Integrable  Heisenberg Ferromagnet Equations  with self-consistent potentials}

\author{ Zh.Kh. Zhunussova,  K.R. Yesmakhanova, D.I. Tungushbaeva, \\ G.K. Mamyrbekova, G.N. Nugmanova and    R. Myrzakulov\footnote{The corresponding author.
Email: rmyrzakulov@gmail.com}
 \\ \textit{Eurasian International Center for Theoretical Physics and  Department of General } \\ \textit{ $\&$  Theoretical Physics, Eurasian National University, Astana 010008, Kazakhstan}}

\date{}
 \maketitle


 \renewcommand{\baselinestretch}{1.1}

 \begin{abstract}
In this paper, we consider some integrable  Heisenberg Ferromagnet Equations  with self-consistent potentials.  We study their Lax representations. In particular we give their equivalent counterparts which are  nonlinear Schr\"odinger type equations.  We present the
integrable reductions of the Heisenberg Ferromagnet Equations  with self-consistent potentials.  These integrable Heisenberg Ferromagnet Equations  with self-consistent
 potentials describe  nonlinear waves in  ferromagnets with some additional physical  fields.
 \end{abstract}

 \tableofcontents
\section{Introduction}
Nonlinear effects are fundamental part of many phenomena in different branches of sciences. Such nonlinear effects are modelled by nonlinear 
differential equations (NDE). One of important parts  of NDE is integrable NDE, which sometimes also called as soliton equations.  Integrable spin systems (SS) are one of main sectors of integrable NDE and play interesting role in mathematics in particular in the geometry of curves and surfaces. On the other hand, integrable SS  play cruical role in the description of nonlinear phenomena in magnets.

In this paper, we study some integrable Myrzakulov equations with self-consistent potentials. We present their Lax representations as well as their reductions. Finally we give their equivalent counterparts which have the nonlinear Schr\"odinger equation type form.

The paper is organized as follows. 
In Sec.\ II, we give  the basic facts from the theory of the Heisenberg ferromagnet equation.  
In Sec.\ III, we investigate the (1+1)-dimensional M-XCIX equation. 
Next, we study the (1+1)-dimensional M-LXIV equation  in Sec.\ IV.  In Sec. V 
we consider the (1+1)-dimensional M-XCIV equation. 
Finally, we give conclusions in Sec.\ VI. 
\section{Preliminaries}

First  example of integrable SS is the so-called  Heisenberg ferromagnetic model (HFM) which reads as  \cite{L1}-\cite{T1}
\begin{equation}
{\bf S}_{t}={\bf S}\wedge {\bf S}_{xx}, \label{HFM}\end{equation}
where $\wedge$ denotes a vector product and \begin{equation}
{\bf S}=(S_1,S_2,S_3), \quad {\bf S}^2=1. \label{2.3} \end{equation}
The matrix form of the HFM looks like 
\begin{equation}
iS_{t}=\frac{1}{2}[S,S_{xx}],\label{2.1} \end{equation} 
 where
 \begin{equation}\label{2.3}
S=S_i\sigma_i=\begin{pmatrix} S_3 &S^{+}\\S^{-} & -S_3\end{pmatrix}.
 \end{equation}
Here $S^2=I, \quad S^{\pm}=S_1\pm iS_2, \quad [A,B]=AB-BA$ and $\sigma_i$ are Pauli matrices
\begin{equation}\label{2.3}
\sigma_1=\begin{pmatrix} 0&1\\1& 0\end{pmatrix}, \quad \sigma_2=\begin{pmatrix} 0&i\\-i& 0\end{pmatrix}, \quad \sigma_3=\begin{pmatrix} 1&0\\0& -1\end{pmatrix}.
 \end{equation}
 Note that the HFM (\ref{HFM}) is Lakshmanan equivalent \cite{L1} to the nonlinear Schr\"odinger equation (NSE)
 \begin{equation}
i\varphi_{t}+\varphi_{xx}+2|\varphi|^2\varphi=0.\label{NSE}
 \end{equation}
Also we recall that between the HFE (\ref{HFM}) and NSE (\ref{NSE}) takes place the gauge equivalence \cite{T1}. 
In literature  different types integrable and nonintegrable SS have been proposed (see e.g. \cite{M0}). As examples of such extensions we here present the following two integrable equations:\\
i) the Myrzakulov-XXXIV (M-XXXIV) equation \cite{M0}
\begin{eqnarray}
{\bf S}_{t}-{\bf S}\wedge {\bf S}_{xx}-u{\bf S}_{x}&=&0,\label{2.1}\\
 u_t+u_x+\alpha({\bf S}_{x}^2)_{x}&=&0.\label{2.2} \end{eqnarray} 
ii) the Myrzakulov-I (M-I) equation \cite{M0}
\begin{eqnarray}
{\bf S}_{t}-({\bf S}\wedge {\bf S}_{y}+u{\bf S})_{x}&=&0,\label{Ia}\\
 u_x+{\bf S}\cdot({\bf S}_{x}\wedge {\bf S}_{y})&=&0.\label{Ib} \end{eqnarray} 
Some properties of these and other integrable and nonintegrable SS were studied in \cite{M}-\cite{C2}. Also  note that the M-I equation (\ref{Ia})-(\ref{Ib}) we write sometimes  as \cite{M0}
\begin{eqnarray}
{\bf S}_{t}-{\bf S}\wedge {\bf S}_{xy}-u{\bf S}_{x}&=&0,\label{Ic}\\
 u_x+{\bf S}\cdot({\bf S}_{x}\wedge {\bf S}_{y})&=&0.\label{Id} \end{eqnarray} 
 Of course that both forms of the M-I equation that is Eq.(\ref{Ia})-(\ref{Ib}) and Eq.(\ref{Ic})-(\ref{Id}) are equivalent each to others. In this paper we study some integrable generalizations of the HFM (2.1).
 

\section{The (1+1)-dimensional M-XCIX equation}
The (1+1)-dimensional Myrzakulov-XCIX equation (or shortly M-XCIX equation) reads as \cite{M0}-\cite{M00}
\begin{eqnarray}
{\bf S}_{t}+0.5\epsilon_1{\bf S}\wedge {\bf S}_{xx}+\frac{2}{\omega}{\bf S}\wedge {\bf W}&=&0,\label{M2a}\\
 {\bf W}_{x}+2\omega {\bf S}\wedge  {\bf W}&=&0,\label{M2b} \end{eqnarray} 
 where $\wedge$ denotes a vector product and \begin{equation}
{\bf S}=(S_1,S_2,S_3), \quad {\bf W}=(W_1,W_2,W_3),\label{2.3} \end{equation} 
  Here $\alpha$ is a real function, ${\bf S}^2=S_{1}^2+S_{2}^2+S_{3}^2=1$, $S_i$ and $W_i$ are some real functions, $\omega$ and $\epsilon_i$ are  real constants.  In terms of components the system (\ref{M2a})-(\ref{M2b}) takes the form
  \begin{eqnarray}
S_{1t}+0.5\epsilon_1( S_{2} S_{3xx}-S_{3}S_{2xx})+\frac{2}{\omega}( S_{2}W_3-S_{3}W_{2})&=&0,\label{2.1}\\
S_{2t}+0.5\epsilon_1( S_{3} S_{1xx}-S_{1}S_{3xx})+\frac{2}{\omega}( S_{3}W_1-S_{1}W_{3})&=&0,\label{2.1}\\
S_{3t}+0.5\epsilon_1( S_{1} S_{2xx}-S_{2}S_{1xx})+\frac{2}{\omega}( S_{1}W_2-S_{2}W_{1})&=&0,\label{2.1}\\
W_{1x}+2\omega (S_{2}W_3-S_3W_2)&=&0,\label{2.2}\\ 
W_{2x}+2\omega (S_{3}W_1-S_1W_3)&=&0,\label{2.2}\\ 
W_{3x}+2\omega (S_{1}W_2-S_2W_1)&=&0.\label{2.2} 
\end{eqnarray}  
On the other hand, the system (\ref{M2a})-(\ref{M2b}) can be rewritten as
\begin{eqnarray}
iS_{t}+0.25\epsilon_1[S, S_{xx}]+\frac{1}{\omega}[S, W]&=&0,\label{2.1}\\
 iW_{x}+\omega [S, W]&=&0,\label{2.2} \end{eqnarray} 
 where
 \begin{equation}\label{2.3}
S=S_i\sigma_i=\begin{pmatrix} S_3 &S^{-}\\S^{+} & -S_3\end{pmatrix}, \quad W=W_i\sigma_i=\begin{pmatrix} W_3 &W^{+}\\W^{-} & -W_3\end{pmatrix}.
 \end{equation}
Here $S^{\pm}=S_1\pm iS_2, \quad W^{\pm}=W_1\pm i W_2,$  $[A,B]=AB-BA, $ $\sigma_i$ are Pauli matrices.
 
 \subsection{Lax representation}
 
Let us consider the system of the  linear equations 
\begin{eqnarray}
\Phi_{x}&=&U\Phi,\label{La}\\
\Phi_{t}&=&V\Phi.\label{Lb} 
\end{eqnarray}  
Let the Lax pair $U-V$ has the form \cite{M0}-\cite{M00}
 \begin{eqnarray}
U&=&-i\lambda S,\label{2.1}\\
V&=&\lambda^2V_2+\lambda V_{1}+\frac{i}{\lambda+\omega}V_{-1}-\frac{i}{\omega}V_{0},\label{2.2} 
\end{eqnarray} 
where
\begin{eqnarray}
V_2&=&-i\epsilon_1 S,\label{2.8}\\
V_1&=&0.25\epsilon_1[S,S_x],\label{2.1}\\
V_{-1}&=&V_0=\begin{pmatrix} W_3&W^{+}\\W^{-}& -W_3\end{pmatrix}.\label{2.2} 
\end{eqnarray} 
With such $U,V$ matrices,  the equation 
\begin{equation}\label{3.21}
U_t-V_x+[U,V]=0
 \end{equation}
is equivalent to the M-XCIX equation (\ref{M2a})-(\ref{M2b}). It means that  the M-XCIX equation (\ref{M2a})-(\ref{M2b}) is integrable by the Inverse Tranform Method (ITM). 

\subsection{Shcr\"odinger-type equivalent counterpart}
Our aim in this section is to find the Shcr\"odinger-type equivalent counterpart of the M-XCIX equation. To do is, let us we introduce the  3 new functions $\varphi$, $p$ and $\eta$ as
\begin{eqnarray}
\varphi&=&\alpha e^{i\beta},\label{2.8}\\
p&=&-\left[2S^{-}W_3-(S_3+1)W^{-}+\frac{S^{-2}W^{+}}{S_3+1}\right]e^{i\varsigma},\label{2.1}\\
\eta&=&2S_3W_3+S^{-}W^{+}+S^{+}W^{-},\label{2.2} 
\end{eqnarray}
where
\begin{eqnarray}
\alpha&=&0.5(S^2_{1x}+S^2_{2x}+S^2_{3x})^{0.5},\label{2.8}\\
\beta&=&-i\partial^{-1}_{x}\left[\frac{tr(S_xSS_{xx})}{tr(S_x^2)}\right],\label{3.1}\\
\varsigma&=&\exp\left[i\theta-\frac{1}{2}\partial_x^{-1}\left(\frac{S^+S^-_x-S^+_xS^-}{1+S_3}\right)\right]\label{2.2} 
\end{eqnarray}
and $\theta=const$.
It is not difficult to verify that these 3 new functions satisfy the following equations\begin{eqnarray}
i\varphi_{t}+\epsilon_1(0.5\varphi_{xx}+|\varphi|^2\varphi)-2ip&=&0, \label{HMBa}\\
p_{x}-2i\omega p -2\eta\varphi&=&0,\label{HMBb}\\
\eta_{x}+\varphi^{*} p +\varphi p^{*}&=&0,\label{HMBc}
 \end{eqnarray}
 It is nothing but the nonlinear Schr\"odinger-Maxwell-Bloch equation (NSMBE). It is well-known that the SMBE is integrable by IST. Its Lax representation reads as \cite{Gabitov}-\cite{Porsezian}
 \begin{eqnarray}
\Psi_{x}&=&A\Psi,\label{La}\\
\Psi_{t}&=&B\Psi,\label{Lb} 
\end{eqnarray}  
where 
 \begin{eqnarray}
A&=&-i\lambda \sigma_3+A_0,\label{3.33}\\
B&=&\lambda^2B_2+\lambda B_{1}+B_0+\frac{i}{\lambda+\omega}B_{-1}.\label{2.2} 
\end{eqnarray} 
Here
\begin{eqnarray}
A_0&=&\begin{pmatrix} 0&\varphi\\-\varphi^{*}& 0\end{pmatrix},\label{2.2}\\
B_2&=&-i\epsilon_1 \sigma_3,\label{2.8}\\
B_1&=&\epsilon_1 A_0,\label{3.37}\\
B_0&=&0.5i\epsilon_1\alpha^2\sigma_3+0.5i\epsilon_1\sigma_3A_{0x},\\
B_{-1}&=&\begin{pmatrix} \eta&-p\\-p^{*}& -\eta\end{pmatrix}.\label{2.2} 
\end{eqnarray} 

\subsection{Reductions}
\subsubsection{Principal chiral equation}

Let us we set $\epsilon_1=0$. Then the M-XCIX equation reduces to the equation
\begin{eqnarray}
iS_{t}+\frac{1}{\omega}[S, W]&=&0,\label{3.40}\\
 iW_{x}+\omega [S, W]&=&0.\label{3.41} \end{eqnarray} 
 It is nothing but the principal chiral equation. As is well-known that it is integrable by ITM. The corresponding Lax pair is given by
 \begin{eqnarray}
U&=&-i\lambda S,\label{3.42}\\
V&=&-\frac{i\lambda}{\omega(\lambda+\omega)}W.\label{3.43} 
\end{eqnarray} 

\subsubsection{Heisenberg ferromagnetic equation}

Now let us we assume that $W=0$. Then the M-XCIX equation reduces to the equation
\begin{equation}
iS_{t}+0.25\epsilon_1[S, S_{xx}]=0.\label{2.2} \end{equation} 
It is the HFM (\ref{HFM}) within to the simplest scale transformations.

\section{The (1+1)-dimensional M-LXIV equation}

The (1+1)-dimensional M-LXIV equation (or shortly M-LXIV equation) reads as \cite{M0}:
\begin{eqnarray}
iS_{t}+\epsilon_2i[ S_{xxx}+6(\beta  S)_{x}]+\frac{1}{\omega}[S, W]&=&0,\label{2.1}\\
 iW_{x}+\omega [S, W]&=&0.\label{2.2} \end{eqnarray} 
 The corresponding  Lax pair is given by
 \begin{eqnarray}
U&=&-i\lambda S,\label{2.1}\\
V&=&\lambda^3V_3+\lambda^2V_2+\lambda V_{1}+\frac{i}{\lambda+\omega}V_{-1}-\frac{i}{\omega}V_{-1},\label{2.2} 
\end{eqnarray} 
where \cite{M0}
\begin{eqnarray}
V_3&=&-4i\epsilon_2 S,\label{2.8}\\
V_2&=&2\epsilon_2SS_x,\label{2.8}\\
V_1&=&\epsilon_2i( S_{xx}+6\beta  S),\label{2.1}\\
V_{-1}&=&W=\begin{pmatrix} W_3&W^{+}\\W^{-}& -W_3\end{pmatrix}\label{2.2} 
\end{eqnarray} 
with $\beta=rq=0.125 tr[(S_x)^2]$. The formulas (3.21)-(3.23) gives us the Schrodinger equivalent of the (1+1)-dimensional M-XCIV equation. It   has the form (see e.g. \cite{C1}-\cite{C2})
\begin{eqnarray}
iq_{t}+i\epsilon_2(q_{xxx}+6rqq_x)-2ip&=&0, \label{z20}\\
ir_{t}+i\epsilon_2(r_{xxx}+6rqr_x)-2ik&=&0, \label{z20}\\
p_{x}-2i\omega p -2\eta q&=&0,\label{z21}\\
k_{x}+2i\omega k -2\eta r&=&0,\label{z21}\\
\eta_{x}+rp +kq&=&0.\label{z22}
 \end{eqnarray}
 This system is nothing but the Hirota-Maxwell-Bloch equation. Its Lax representation reads as
 \begin{eqnarray}
\Psi_{x}&=&A\Psi,\label{z7}\\
\Psi_{t}&=&[-4i\epsilon_2\lambda^3\sigma_3+B]\Psi,\label{z8} 
\end{eqnarray}  
where 
 \begin{eqnarray}
A&=&-i\lambda \sigma_3+A_0,\label{z9}\\
B&=&\lambda^2B_2+\lambda B_1+B_0+\frac{i}{\lambda+\omega}B_{-1}.\label{z10} 
\end{eqnarray} 
Here
\begin{eqnarray}
B_2&=&4\epsilon_2A_{0},\label{z11}\\
B_1&=&2i\epsilon_2rq\sigma_3+2i\epsilon_2\sigma_3A_{0x},\label{z11}\\
A_0&=&\begin{pmatrix} 0&q\\-r& 0\end{pmatrix},\label{z12}\\
B_0&=&\epsilon_2(r_{x}q-rq_{x})\sigma_3+B_{01}, \\
B_{01}&=&\begin{pmatrix} 0&-\epsilon_2q_{xx}-2\epsilon_2rq^2\\\epsilon_2r_{xx}+2\epsilon_2qr^2& 0\end{pmatrix},\label{z13}\\
B_{-1}&=&\begin{pmatrix} \eta&-p\\-k& -\eta\end{pmatrix}.\label{z14} 
\end{eqnarray}
This system we can reduce to the form
\begin{eqnarray}
iq_{t}+i\epsilon_2(q_{xxx}+6\delta|q|^2q_x)-2ip&=&0, \label{z20}\\
p_{x}-2i\omega p -2\eta q&=&0,\label{z21}\\
\eta_{x}+\delta(q^{*} p +p^{*} q)&=&0.\label{z22}
 \end{eqnarray}

\section{The (1+1)-dimensional M-XCIV equation}

The Myrzakulov-XCIV equation or shortly M-XCIV equation reads as \cite{M0}:
\begin{eqnarray}
iS_{t}+0.5\epsilon_1[S, S_{xx}]+\epsilon_2i[ S_{xxx}+6(\beta  S)_{x}]+\frac{1}{\omega}[S, W]&=&0,\label{2.1}\\
 iW_{x}+\omega [S, W]&=&0.\label{2.2} \end{eqnarray} 
 \subsection{Lax representation} The  Lax pair of the M-XCIV equation (5.1)-(5.2) is given by
 \begin{eqnarray}
U&=&-i\lambda S,\label{2.1}\\
V&=&\lambda^3V_3+\lambda^2V_2+\lambda V_{1}+\frac{i}{\lambda+\omega}V_{-1}-\frac{i}{\omega}V_{-1},\label{2.2} 
\end{eqnarray} 
where \cite{M0}
\begin{eqnarray}
V_3&=&-4i\epsilon_2 S,\label{2.8}\\
V_2&=&-2i\epsilon_1 S+2\epsilon_2SS_x,\label{2.8}\\
V_1&=&\epsilon_1SS_x+\epsilon_2i( S_{xx}+6\beta  S),\label{2.1}\\
V_{-1}&=&W=\begin{pmatrix} W_3&W^{+}\\W^{-}& -W_3\end{pmatrix}\label{2.2} 
\end{eqnarray} 
with $\beta=rq=0.125 tr[(S_x)^2]$. 
 \subsection{Reductions}
 The M-XCIV equation  admits some  integrable reductions. For example, it has the following integrable reductions.
 
 \subsubsection{The M-XCIX equation}
 Let $\epsilon_2=0$. Then the M-XCIV equation takes the form
 \begin{eqnarray}
iS_{t}+0.5\epsilon_1[S, S_{xx}]+\frac{1}{\omega}[S, W]&=&0,\label{2.1}\\
 iW_{x}+\omega [S, W]&=&0.\label{2.2} \end{eqnarray}
 It has the   Lax pair of the form
 \begin{eqnarray}
U&=&-i\lambda S,\label{2.1}\\
V&=&\lambda^3V_3+\lambda^2V_2+\lambda V_{1}+\frac{i}{\lambda+\omega}W-\frac{i}{\omega}W,\label{2.2} 
\end{eqnarray} 
where \cite{M0}
\begin{eqnarray}
V_2&=&-2i\epsilon_1 S,\label{2.8}\\
V_1&=&\epsilon_1SS_x,\label{2.1}\\
W&=&\begin{pmatrix} W_3&W^{+}\\W^{-}& -W_3\end{pmatrix}.\label{2.2} 
\end{eqnarray} 
 
 \subsubsection{The M-LXIV equation}
 Now let us consider the case $\epsilon_1=0$. In this case the M-XCIV equation transforms to the equation
 \begin{eqnarray}
iS_{t}+\epsilon_2i[ S_{xxx}+6(\beta  S)_{x}]+\frac{1}{\omega}[S, W]&=&0,\label{2.1}\\
 iW_{x}+\omega [S, W]&=&0.\label{2.2} \end{eqnarray}
 The corresponding Lax pair reads as
  \begin{eqnarray}
U&=&-i\lambda S,\label{2.1}\\
V&=&\lambda^3V_3+\lambda^2V_2+\lambda V_{1}+\frac{i}{\lambda+\omega}V_{-1}-\frac{i}{\omega}V_{-1},\label{2.2} 
\end{eqnarray} 
where \cite{M0}
\begin{eqnarray}
V_3&=&-4i\epsilon_2 S,\label{2.8}\\
V_2&=&2\epsilon_2SS_x,\label{2.8}\\
V_1&=&\epsilon_2i( S_{xx}+6\beta  S),\label{2.1}\\
V_{-1}&=&W=\begin{pmatrix} W_3&W^{+}\\W^{-}& -W_3\end{pmatrix}\label{2.2} 
\end{eqnarray} 
with $\beta=rq=0.125 tr[(S_x)^2]$. 
\subsection{Equivalent counterpart}To find the Schrodinger equivalent, we again us the formulas (3.21)-(3.23). Finally the Schrodinger equivalent of the (1+1)-dimensional M-XCIV equation  has the form (see e.g. \cite{C1}-\cite{C2})
\begin{eqnarray}
iq_{t}+\epsilon_1(q_{xx}+2rq^2)+i\epsilon_2(q_{xxx}+6rqq_x)-2ip&=&0, \label{z20}\\
ir_{t}-\epsilon_1(r_{xx}+2r^2q)+i\epsilon_2(r_{xxx}+6rqr_x)-2ik&=&0, \label{z20}\\
p_{x}-2i\omega p -2\eta q&=&0,\label{z21}\\
k_{x}+2i\omega k -2\eta r&=&0,\label{z21}\\
\eta_{x}+rp +kq&=&0.\label{z22}
 \end{eqnarray}
 This system is nothing but the Hirota-Maxwell-Bloch equation. Its Lax representation reads as
 \begin{eqnarray}
\Psi_{x}&=&A\Psi,\label{z7}\\
\Psi_{t}&=&[-4i\epsilon_2\lambda^3\sigma_3+B]\Psi,\label{z8} 
\end{eqnarray}  
where 
 \begin{eqnarray}
A&=&-i\lambda \sigma_3+A_0,\label{z9}\\
B&=&\lambda^2B_2+\lambda B_1+B_0+\frac{i}{\lambda+\omega}B_{-1}.\label{z10} 
\end{eqnarray} 
Here
\begin{eqnarray}
B_2&=&-2i\epsilon_1\sigma_3+4\epsilon_2A_{0},\label{z11}\\
B_1&=&2i\epsilon_2rq\sigma_3+2i\epsilon_2\sigma_3A_{0x}+2\epsilon_1A_0,\label{z11}\\
A_0&=&\begin{pmatrix} 0&q\\-r& 0\end{pmatrix},\label{z12}\\
B_0&=&(i\epsilon_1rq+\epsilon_2(r_{x}q-rq_{x}))\sigma_3+B_{01}, \\
B_{01}&=&\begin{pmatrix} 0&i\epsilon_1q_x-\epsilon_2q_{xx}-2\epsilon_2rq^2\\i\epsilon_1r_x+\epsilon_2r_{xx}+2\epsilon_2qr^2& 0\end{pmatrix},\label{z13}\\
B_{-1}&=&\begin{pmatrix} \eta&-p\\-k& -\eta\end{pmatrix}.\label{z14} 
\end{eqnarray}
If $p=\delta k^{*}, r=\delta q^{*}$, this system we can reduce to the form
\begin{eqnarray}
iq_{t}+\epsilon_1(q_{xx}+2\delta|q|^2q)+i\epsilon_2(q_{xxx}+6\delta|q|^2q_x)-2ip&=&0, \label{z20}\\
p_{x}-2i\omega p -2\eta q&=&0,\label{z21}\\
\eta_{x}+\delta(q^{*} p +p^{*} q)&=&0.\label{z22}
 \end{eqnarray}
 
 Note that the (1+1)-dimensional HMBE (\ref{z20})-(\ref{z22}) admits the following integrable  reductions.
 
 i) The NSLE as $\epsilon_1-1=\epsilon_2=p=\eta=0$:
 \begin{eqnarray}
iq_{t}+q_{xx}+2\delta|q|^2q=0.\label{z23}
 \end{eqnarray}
 
 ii) The (1+1)-dimensional complex mKdV eqation as $\epsilon_1=\epsilon_2-1=p=\eta=0$:
 \begin{eqnarray}
q_{t}+q_{xxx}+6\delta|q|^2q_x=0. \label{z24}\
 \end{eqnarray}
 
 iii) The (1+1)-dimensional Schrodinger-Maxwell-Bloch equation as $\epsilon_1-1=\epsilon_2=0$:
 \begin{eqnarray}
iq_{t}+q_{xx}+2\delta|q|^2q-2ip&=&0, \label{z20}\\
p_{x}-2i\omega p -2\eta q&=&0,\label{z21}\\
\eta_{x}+\delta(q^{*} p +p^{*} q)&=&0.\label{z22}
 \end{eqnarray}
 
 iv) The (1+1)-dimensional complex mKdV-Maxwell-Bloch  equation as $\epsilon_1=\epsilon_2-1=0$:
  \begin{eqnarray}
q_{t}+q_{xxx}+6\delta|q|^2q_x-2p&=&0, \label{z20}\\
p_{x}-2i\omega p -2\eta q&=&0,\label{z21}\\
\eta_{x}+\delta(q^{*} p +p^{*} q)&=&0.\label{z22}
 \end{eqnarray}
 
 v) The following (1+1)-dimensional  equation as $\epsilon_1=\epsilon_2=0$:
  \begin{eqnarray}
q_{t}-2p&=&0, \label{z20}\\
p_{x}-2i\omega p -2\eta q&=&0,\label{z21}\\
\eta_{x}+\delta(q^{*} p +p^{*} q)&=&0.\label{z22}
 \end{eqnarray}
 or
  \begin{eqnarray}
q_{xt}-2i\omega q_t -4\eta q&=&0,\label{z21}\\
2\eta_{x}+\delta(|q|^{2})_t&=&0.\label{z22}
 \end{eqnarray}
 
 vi) The following (1+1)-dimensional  equation as $\delta=0$:
 \begin{eqnarray}
iq_{t}+\epsilon_1q_{xx}+i\epsilon_2q_{xxx}-2ip&=&0, \label{z20}\\
p_{x}-2i\omega p -2\eta_0 q&=&0,\label{z22}
 \end{eqnarray}
 where $\eta_0=0$. Again we note that all these reductions are integrable by IST. The corresponding Lax representations we get from the Lax representation (5.29)-(5.30) as the corresponding reductions.
\section{Conclusion}
Heisenberg ferromagnet models play an important role in modern theory of magnets. These are nonlinear partial  differential equations. Some of  these models  are integrable by the Inverse Scattaring Method that is they are soliton equations.  In this paper, we have studied some Heisenberg ferromagnet equations (models) with self-consistent potentials. We have presented their Lax representations. Also we have found their Schr\"odinger type equivalent counterparts.

\end{document}